\documentclass[aps,prl,twocolumn,showpacs,superscriptaddress]{revtex4-1}
\usepackage{graphicx}
\usepackage{MnSymbol}
\usepackage{bm}  

\newcommand  {\at}   {\mathrm{at}}

\newcommand  {\sat}  {\mathrm{sat}}

\newcommand {\INLN} {Universit\'e C\^ote d'Azur, CNRS, INLN, 06560 Valbonne, France}
\newcommand {\Brazil} {CAPES Foundation, Ministry of Education of Brazil, Bras\'{i}lia, DF
70040-020, Brazil}
\newcommand {\Strathclyde} {SUPA and Department of Physics, University of Strathclyde, Glasgow G4 0NG, Scotland, United Kingdom}

\begin{document}

\title{Superradiance in a Large and Dilute Cloud of Cold Atoms in the Linear-Optics Regime}

\author{Michelle O. Ara\'{u}jo}
\affiliation{\INLN}
\affiliation{\Brazil}
\author{Ivor Kre\v si\' c}
\affiliation{\Strathclyde}
\author{Robin Kaiser}
\affiliation{\INLN}
\author{William Guerin}
\email{william.guerin@inln.cnrs.fr}
\affiliation{\INLN}


\date{\today}

\begin{abstract}
Superradiance has been extensively studied in the 1970s and 1980s in the regime of superfluorescence, where a large number of atoms are initially excited.
Cooperative scattering in the linear-optics regime, or ``single-photon superradiance'', has been investigated much more recently, and superradiant decay has also been predicted, even for a spherical sample of large extent and low density, where the distance between atoms is much larger than the wavelength.
Here, we demonstrate this effect experimentally by directly measuring the decay rate of the off-axis fluorescence of a large and dilute cloud of cold rubidium atoms after the sudden switch-off of a low-intensity laser driving the atomic transition. We show that, at large detuning, the decay rate increases with the on-resonance optical depth. In contrast to forward scattering, the superradiant decay of off-axis fluorescence is suppressed near resonance due to attenuation and multiple-scattering effects.
\end{abstract}


\maketitle

In his classic paper on coherence in spontaneous radiation by atomic samples~\cite{Dicke:1954}, Dicke showed that a collection of identical excited atoms could synchronize to emit light coherently. In the case initially envisioned by Dicke, an atomic sample of size small compared to the wavelength of the transition, superradiance can be interpreted as the spontaneous synchronization of the radiation by all atoms and can be understood by the buildup of a giant dipole corresponding to the symmetric superposition of atomic states. Since it is difficult to prepare such dense and small samples, and since near-field dipole-dipole interaction may in fact prevent superradiance at high density~\cite{Friedberg:1974}, experimental studies of superradiance in the 1970s and 1980s have been realized with large-size samples (mainly pencil-shaped) of low density~\cite{Feld:1980,Gross:1982}. In this regime, superradiance, more precisely called superfluorescence~\cite{Bonifacio:1975,Malcuit:1987}, is intrinsically a nonlinear optical process.

More recently, it has been pointed out that a single photon, first absorbed by one atom among $N$ others in a sample of large size and low density, would be spontaneously emitted in the direction of the initial photon wave vector~\cite{Scully:2006}, in contrast with the simple picture of isotropic spontaneous emission. This coherent forward scattering, which has been observed very recently~\cite{Bromley:2016}, can be explained by a phase-matching condition, and thus does not rely on dipole-dipole interactions. This extended-volume regime was already mentioned by Dicke~\cite{Dicke:1954} and was further developed by others~\cite{Arecchi:1970,Rehler:1971}.

A less obvious result, which does rely on the long-range, light-induced dipole-dipole interactions between atoms, is the decay rate $\Gamma_N$ of the corresponding collective excited state, which has been computed by many authors~\cite{Arecchi:1970,Rehler:1971,Mazets:2007,Svidzinsky:2008,Svidzinsky:2008b,Courteille:2010,Friedberg:2010,Prasad:2010,Skipetrov:2011},
\begin{equation}\label{eq.GammaN}
\Gamma_N = \mathcal{C} \frac{N}{(kR)^2} \Gamma \; ,
\end{equation}
where $\Gamma^{-1}$ is the lifetime of the excited state of a single-atom in vacuum, $N$ is the number of atoms, $k=2\pi/\lambda$ is the wave vector associated with the optical transition, $R$ is the radius of the sample, and $\mathcal{C}$ is a numerical factor on the order of unity, which depends on the precise geometry of the sample. If the number of atoms is sufficiently large, one can have $\Gamma_N \gg \Gamma$, corresponding to a superradiant decay, even at low spatial density, where the separation between atoms is much larger than the wavelength. This is in contrast with the case of two particles~\cite{DeVoe:1996,Barnes:2005,McGuyer:2015}, for which the single-atom decay rate is recovered for an atomic separation larger than $\lambda$. This ``single-photon superradiance'' has attracted a lot of attention in the last years~\cite{Scully:2009,Manassah:2012,Bienaime:2013}, but direct experimental evidence has been limited to special geometries involving cavities or waveguides~\cite{Roehlsberger:2010,Goban:2015} or to multilevel schemes~\cite{Srivathsan:2013,Oliveira:2014}. Related phenomena are optical precursors~\cite{Jeong:2006,Chen:2010} or ``flash''~\cite{Chalony:2011,Kwong:2014,Kwong:2015}, which can also have a temporal dynamic faster than $\Gamma$. Since these effects come from the interference between the scattered field and the driving field, they are only visible in the forward direction and can be explained by the transient response of the complex refractive index of the gas.

\begin{figure*}[t]
\centering
\includegraphics{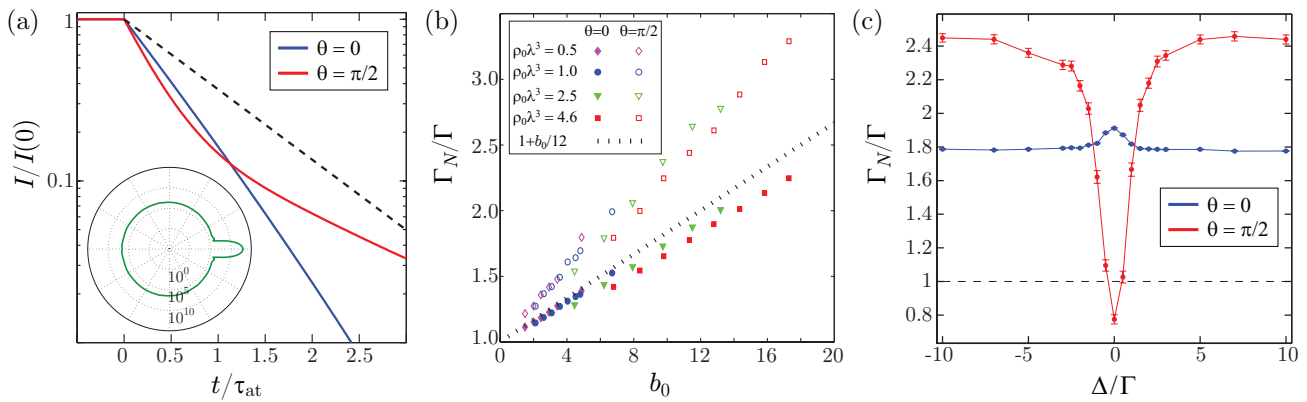}
\caption{Numerical study of the initial collective decay rate $\Gamma_N$. (a) Temporal evolution of the fluorescence after the switch off of the driving laser at $t=0$, with $b_0=11.3$, $\rho_0 \lambda^3 = 4.6$, $\Delta=10\Gamma$, averaged over 50 configurations, for two different angles, in the forward direction $\theta=0$ and at $90^\circ$ ($\theta = \pi/2$). The amplitude is normalized to the steady state amplitude, which is much larger for $\theta=0$ as shown in the emission diagram (inset, in log scale). The time axis is normalized to the lifetime of the excited state $\tau_\at = \Gamma^{-1} \approx 26$~ns. An exponential fit in the range $0<t/\tau_\at\leq0.2$ allows us to extract the initial decay rate $\Gamma_N$. At late time, the decay becomes subradiant~\cite{Bienaime:2012,Guerin:2016}. The dashed line shows the decay expected for a single atom (rate $\Gamma$). (b) Decay rate as a function of the resonant optical thickness $b_0 = 3N/(kR)^2$ for different densities ($\rho_0$ is the density at the center of the cloud). Filled symbols are for $\theta=0$ and open symbols for $\theta=\pi/2$. The increase is mainly linear in $b_0$. The slope of the linear increase slightly depends on the angle. The dotted line shows the expectation for the decay of the timed-Dicke state [Eq.~(\ref{eq.TD})]. (c) Decay rate as a function of the detuning, for $b_0=17$, $\rho_0 \lambda^3 = 4.6$ and detection angles $\theta = 0, \pi/2$. Off-axis superradiance is suppressed near resonance. The error bars shown in panel (c) and omitted in panel (b) for clarity correspond to the 95\% confidence interval for the exponential fit of the decay rate.}
\label{fig.theory}
\end{figure*}

On the contrary, light emission at different angles (``off-axis scattering'' or ``fluorescence'') cannot be explained by a phase-matching condition imposed by the initial laser field or a continuous-medium description~\cite{Bachelard:2012}. Recently, we have used off-axis scattering to observe subradiance~\cite{Bienaime:2012,Guerin:2016}. In this Letter, we present the direct observation of the superradiant decay of the fluorescence emitted in free space by a large spherical sample of cold atoms, which is continuously driven by a low-intensity laser that is abruptly switched off.

A true single-photon source is indeed not necessary to observe single-photon superradiance. As stressed by Prasad and Glauber~\cite{Prasad:2010}, it is the spatially extended initial coherence, not entanglement \textit{per se}, that is fundamentally responsible for cooperative radiation processes such as superradiance and subradiance (see also Refs.~\cite{Eberly:2006,Friedberg:2007,Bienaime:2011}), so that continuous driving by a low-intensity laser (compared to the saturation intensity of the atomic transition) can also be used to study these effects~\cite{Courteille:2010,Friedberg:2010}. Similarly, it has been shown that the full quantum problem is equivalent, in the linear-optics regime, to classical coupled dipoles~\cite{Javanainen:1999,Svidzinsky:2010}.

Before turning to the experimental results, we use the coupled-dipole model to illustrate the qualitative differences between forward and off-axis scattering.
We consider a sample of $N$ motionless two-level atoms distributed over a 3D Gaussian atomic density distribution of rms radius $R$, illuminated along the $z$ axis by a plane wave (wave vector $\bm{k}_0 = k \bm{\hat{z}}$) with detuning $\Delta = \omega-\omega_0$ and Rabi frequency $\Omega \ll \Gamma$. In the low-intensity limit, using the Markov approximation, the linear response of this many-body system can be simulated by $N$ coupled-dipole equations~\cite{Courteille:2010,Bienaime:2011}
\begin{equation}
\dot{\beta}_i = \left( i\Delta-\frac{\Gamma}{2} \right)\beta_i -\dfrac{i\Omega}{2}e^{i\bm{k}_0 \cdot \bm{r}_i } + \frac{i\Gamma}{2} \sum_{i \neq j} V_{ij}\beta_j \; ,
\label{eq.betas}
\end{equation}
where $\beta_i$ is the amplitude of the single-excited-atom state $|i\rangle = |g \cdots e_i \cdots g\rangle$ and
\begin{equation}
V_{ij} = \frac{e^{ikr_{ij}}}{kr_{ij}} \; , \quad r_{ij} = |\bm{r}_i - \bm{r}_j| \; ,\label{eq.Vij}
\end{equation}
describes the light-induced dipole-dipole interaction in the scalar approximation, neglecting near-field terms and polarization effects, which is a good approximation for our dilute samples~\cite{Skipetrov:2014,Bellando:2014}. The first term of the rhs. of Eq.~(\ref{eq.betas}) corresponds to the natural evolution of the dipoles (oscillation and damping), the second one corresponds to the driving by the external laser, and the last one, the dipole-dipole interaction term, is responsible for cooperative effects.

Numerically solving these equations allows us to compute the emission diagram~\cite{Bienaime:2011} as well as the temporal decay after switching off the driving term~\cite{Bienaime:2012,Guerin:2016}. By fitting the initial decay just after the switch off, we can study how the collective decay rate $\Gamma_N$ depends on the emission direction, on the resonant optical depth $b_0=3N/(kR)^2$~\cite{footnote_b0} and on the detuning $\Delta$. Note that the atom number is limited to a few thousands in the simulations and that the complex Zeeman structure of rubidium atoms is not taken into account, so that a quantitative agreement with the experiment is not expected.

The main results of the numerical study are reported in Fig.~\ref{fig.theory}. \emph{At large detuning}, the steady state reached before switch off tends to the ``driven timed-Dicke state''~\cite{Scully:2006,Courteille:2010,Bienaime:2013}, in which all atoms have the same excitation probability. As for a collection of independent atoms, the emission diagram is mainly forward directed for large $b_0$ [inset of Fig.~\ref{fig.theory}(a)], but it also contains a non-negligible quasi-isotropic background, which is neglected in the continuous-medium approach used in Refs.~\cite{Mazets:2007,Svidzinsky:2008,Svidzinsky:2008b,Courteille:2010,Friedberg:2010,Prasad:2010}. It has also been shown in Ref.~\cite{Bienaime:2011} that this collective state decays with an initial rate
\begin{equation}\label{eq.TD}
\Gamma_N = \left(1+\frac{b_0}{12} \right) \Gamma
\end{equation}
for a Gaussian atomic distribution, which is consistent with the scaling of Eq.~(\ref{eq.GammaN}) for very large $b_0$ and with the single-atom limit for small $b_0$. We observe this scaling in Fig.~\ref{fig.theory}(b). The slope for the forward scattered light ($\theta = 0$) is very close to the one predicted by Eq.~(\ref{eq.TD}) because forward scattering is the most important contribution.
Moreover, maybe surprisingly, the light scattered off axis exhibits superradiant decay as well~\cite{note_skipetrov}, with a similar linear scaling with $b_0$, the slope being slightly modified by the angle difference [Figs.~\ref{fig.theory}(a) and \ref{fig.theory}(b)]. Superradiance is thus also visible, and even with a faster decay rate, in the off-axis scattering.

In an experiment, it is hard to use a very large detuning because it obviously decreases the amount of fluorescence. In practice, using a moderate detuning contributes to populate other states than the timed-Dicke state~\cite{Bienaime:2013}, essentially because of the exponential attenuation of the driving field inside the cloud. This contributes to populate longer-lived states, which can be interpreted as subradiant states at large or moderate detuning~\cite{Guerin:2016} or simply as radiation trapping due to multiple scattering near resonance~\cite{Labeyrie:2003}. We thus expect that the superradiant decay is suppressed near resonance. This is what we observe in the numerical results of Fig.~\ref{fig.theory}(c) for off-axis scattering. The behavior of forward scattering is different because it is related to the transient response of the refractive index. As shown in~\cite{Kwong:2015}, it is slightly faster on resonance. These qualitatively different behaviors of forward and off-axis scattering emphasize that the two are different physical mechanisms. Although this is almost never stated clearly, the forward lobe seen in Fig.~\ref{fig.theory}(a) and discussed in many papers (see, e.g., Refs.~\cite{Svidzinsky:2008,Courteille:2010,Chabe:2014,Bromley:2016}) should indeed be seen as diffracted and refracted light more than scattered light.

\begin{figure}[t]
\centering
\includegraphics{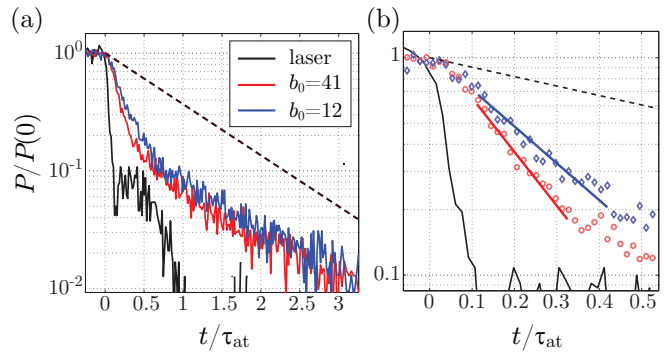}
\caption{(a) Decay of the measured fluorescence power $P$ after switching off the probe laser for two different $b_0$ (red and blue solid lines) at a given detuning $\Delta=-6\Gamma$. The vertical axis is normalized to the steady-state fluorescence level. The dashed line shows the expected decay for a single atom, $e^{-t/\tau_\at}$, and the black solid line is the switch off of the laser (the fast part with a poor extinction ratio is due to the EOM and the slower part is due to the AOM). (b) Same data  at shorter time scales, with the exponential fit of the initial decay that allows us to measure $\Gamma_N$.}
\label{fig.decay}
\end{figure}

Let us now turn to our experimental observation of superradiance.
In our experiment, we load $N\approx 10^9$ $^{87}$Rb atoms from a background vapor into a magneto-optical trap (MOT) for 50~ms. A compressed MOT (30~ms) period allows for an increased and smooth spatial density with a Gaussian distribution of rms radius $R\approx$ 1~mm (typical density $\rho \approx 10^{11}$~cm$^{-3}$) and a reduced temperature $T\approx 50~\mu$K. We then switch off the MOT trapping beams and magnetic field gradient and allow for 3~ms of free expansion, used to optically pump all atoms into the upper hyperfine ground state $F=2$. Next, we apply a series of 12 pulses of a weak probe beam (waist $w = 5.7$~mm), linearly polarized and detuned by $\Delta$ from the closed atomic transition $F=2 \rightarrow F'=3$ ($\lambda=780.24$~nm and $\Gamma/2\pi = 6.07$~MHz). When we varied the detuning, we also varied the laser intensity accordingly in order to keep the saturation parameter approximately constant at $s \simeq (2.2\pm 0.6)\times 10^{-2}$. The pulses of duration $30~\mu$s and separated by $1$~ms are obtained by an electro-optical modulator (EOM, fibered Mach-Zehnder intensity modulator by EOspace, Ref. AZ-0K5-10-PFU-SFU-780) with a 90\%-10\% falltime of about 3~ns (Fig.~\ref{fig.decay}). It is driven by a pulse generator (DG535 by SRS) and actively locked to avoid any drift of the working point. In order to improve the extinction ratio, we also use an acousto-optical modulator (AOM) in series with the EOM. Between subsequent pulses of each series, the size of the cloud increases because of thermal expansion, and the atom number decreases due to off resonant optical pumping into the $F=1$ hyperfine state during each pulse, which allows us to realize different optical depths within one series of pulses. After this stage of expansion and measurement, the MOT is switched on again and most of the atoms are recaptured. The complete cycle is thus short enough to allow the signal integration over a large number of cycles, typically $\sim 500 000$. The fluorescence is collected by a lens with a solid angle of $\sim 5\times 10^{-2}$~sr at  $\theta\approx35^\circ$ from the incident direction of the laser beam and detected by a photomultiplier (Hamamatsu HPM R10467U-50). The signal is then recorded on a multichannel scaler (MCS6A by FAST ComTec) with a time bin of $0.4$~ns, averaging over the cycles. The cooperativity parameter $b_0$ corresponding to each pulse is calibrated by an independent measurement of the atom number, cloud size and temperature using absorption imaging (see the Supplemental Material of Ref.~\cite{Guerin:2016}).

\begin{figure*}[t]
\centering
\includegraphics{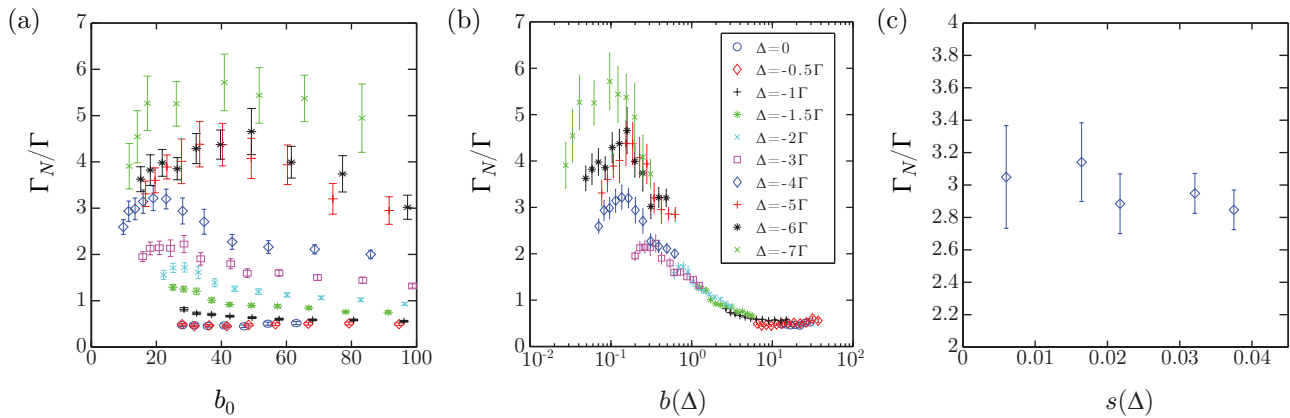}
\caption{Experimental study of the initial collective decay rate $\Gamma_N$. (a) Systematic analysis of $\Gamma_N$ as a function of the resonance optical thickness $b_0$ and the detuning $\Delta$. (b) Same data shown as a function of $b(\Delta)$. When $b(\Delta)\gtrsim 1$, it becomes the scaling parameter. (c) $\Gamma_N$ as a function of the saturation parameter $s(\Delta)$, for $b_0 = 21 \pm 1$ and $\Delta = -4\Gamma$. In all panels, error bars represent the 95\% confidence interval of the fit.}
\label{fig.Gamma_exp}
\end{figure*}

We show in Fig.~\ref{fig.decay} examples of the measured fluorescence decay for different values of $b_0$ and a fixed detuning $\Delta=-6\Gamma$. We clearly see that the decay is much faster than the single-atom decay, in contrast to the behavior of collective \emph{incoherent} scattering effects such as radiation trapping~\cite{Labeyrie:2003}. This fast decay rate increases with $b_0$, in line with the expected superradiant behavior. From these data we can fit the initial decay by an exponential and extract the collective decay rate.
The fitting procedure has been chosen as follows. The range of the fit starts at $t/\tau_\at = 0.1$, when the probe laser intensity has decayed to 10\% of its initial value. It ends when the measured signal arrives at 20\% of its initial value or when the background light scattered from the hot Rb vapor in the vacuum chamber is at this level. This background light decays like $e^{-t/\tau_\at}$ [well visible in the Fig.~\ref{fig.decay} for $P/P(0)<10^{-1}$] and has a relative weight that depends on the atom number and the detuning (it is negligible on resonance with the cold atoms and becomes important far from resonance) and which is independently calibrated for each measurement. Finally, when the number of points in the fitting range is less than 15, or when the statistical coefficient of determination of the fit $R^2$ is less than 0.85, the data are discarded.

The systematic study of the collective decay rate $\Gamma_N$ as a function of the resonant optical thickness $b_0$ and the detuning $\Delta$ is presented in Fig.~\ref{fig.Gamma_exp}(a). The increase with $b_0$ is well visible, especially at large detuning, up to a maximum value $\Gamma_\mathrm{max} \sim 5\Gamma -- 6\Gamma$, well above the decay rate of independent atoms. We note that the curves acquired for different detunings do not collapse on a single curve, contrary to what has been observed for subradiance~\cite{Guerin:2016}, showing the sensitivity of superradiance to the proximity of the resonance. Indeed, the decay rates measured for small detunings do not exhibit superradiance, and even at moderate detuning, the decay rate starts to decrease at high $b_0$, when the actual optical thickness
\begin{equation}
b(\Delta) = g \frac{b_0}{1+4\Delta^2/\Gamma^2} \,
\end{equation}
is on the order of $1$  or higher (here $g=7/15$ is the relative strength of the transition for a statistical mixture of Zeeman states). We show indeed in Fig.~\ref{fig.Gamma_exp}(b) that $b(\Delta)$ becomes the relevant parameter in this regime. These observations are perfectly consistent with the expectation of the coupled-dipole model [Fig.~\ref{fig.theory}(c)] and with the intuition that collective superradiant states are less populated if the driving field is attenuated inside the sample~\cite{Bienaime:2013}.

Finally, we checked that the results are independent of the intensity (or the saturation parameter) to confirm that the experiments have been done in the linear-optics regime. For this we varied the intensity $I$ of the probe beam at fixed detuning and $b_0$, and we report in Fig.~\ref{fig.Gamma_exp}(c) the decay rate as a function of the saturation parameter
\begin{equation}
s(\Delta) = g \frac{I/I_\sat} {1+4\Delta^2/\Gamma^2}\, ,
\end{equation}
with $I_\sat=1.6$~mW/cm$^2$ the saturation intensity. We observe no significant variation of $\Gamma_N$ with the saturation parameter in the explored range $s<0.04$.

In summary, we have reported the first observation of superradiant decay in free space (without a cavity) in the low-intensity regime, using the fluorescence (off-axis scattering) of a cold-atom cloud. We have shown that at large detuning, the decay rate increases with the resonant optical depth, but it is suppressed near resonance. These observations are consistent with numerical solutions of coupled-dipole equations in the dilute limit. The shortening of the radiative lifetime due to cooperativity is potentially important to a number of areas, such as the diagnostics of ultracold gases~\cite{Chomaz:2012,Bons:2016}, quantum memories~\cite{Walther:2009,Oliveira:2014}, optical clocks~\cite{Chang:2004,Bromley:2016,Okaba:2014}, ultranarrow lasers~\cite{Bohnet:2012}, photon-pair sources~\cite{Srivathsan:2013}, and light-harvesting systems~\cite{Monshouwer:1997,Celardo:2014}.

To conclude, let us notice that in Dicke superfluorescence~\cite{Feld:1980}, in optical precursors (or flash)~\cite{Kwong:2015}, as well as in the experiment reported here dealing with low-saturation fluorescence, the time scales associated with the transient regimes are always governed by the same cooperativity parameter, the resonant optical depth. These three phenomena can be related to stimulated emission, the refractive index and spontaneous emission, respectively, and are thus different facets of light-atom interaction. It is interesting, and also beautiful, to see that they exhibit cooperativity in a similar way. On the other hand, other collective properties, such as the cw susceptibility (including the refractive index, the linear attenuation coefficient or gain coefficient for inverted systems, the Lorentz-Lorenz shift~\cite{Friedberg:1973}, and beyond mean-field corrections~\cite{Javanainen:2016,Jennewein:2016}) and weak-localization corrections to diffusive transport~\cite{Rossum:1999}, are governed by the atomic density. The fundamental difference is that the latter are properties of the bulk material, which can be defined for an infinite medium. Transient phenomena, on the contrary, involve light escaping from the sample, in which case the finite size of the medium and the finite number of atoms are necessarily key parameters~\cite{Guerin:JMO}.


We thank R. Bachelard and N. Piovella for fruitful discussions. We acknowledge financial support from the CNRS international program, the French Agence National pour la Recherche (Project LOVE, No. ANR-14-CE26-0032), the Brazilian Coordena\c{c}\~{a}o de Aperfei\c{c}oamento de Pessoal de N\'{i}vel Superior (CAPES), the Brazilian Conselho Nacional de Desenvolvimento Cient\'ifico e Tecnol\'ogico (CNPq, Project PVE, No. 303426/2014-4), and the European Research Executive Agency (Program COSCALI, No. PIRSES-GA-2010-268717).

\ \\

\noindent \textit{Note added.}---Recently, we have learned of a complementary experiment on superradiance in for forward direction; see Ref.~\cite{Havey}.



\begin{thebibliography}{10}
\bibitem{footnote_b0}
We use here the definition of $b_0$ corresponding to a vectorial model for
  light, even if the actual optical thickness in the scalar model is given by
  $2N/(kR)^2=2b_0/3$.
\end{thebibliography}


\end{document}